\documentstyle[aps,psfig,amssymb]{revtex}\begin{document}
\twocolumn\wideabs{
\title{Nearly ideal binary communication in squeezed channels}
\author{Matteo G. A. Paris}\address{Quantum Optics $\&$ Information Group, 
Istituto Nazionale per la Fisica della Materia \\ 
Universit\`a di Pavia, via Bassi 6, I-27100 Pavia, Italy}
\date{\today}\maketitle
\begin{abstract} We analyze the effect of squeezing the channel in binary 
communication based on Gaussian states. We show that for coding on pure 
states, squeezing increases the detection probability at fixed size of the 
strategy, actually saturating the optimal bound already for moderate signal 
energy. Using Neyman-Pearson lemma for fuzzy hypothesis testing we are able 
to analyze also the case of mixed states, and to find the optimal amount 
of squeezing that can be effectively employed. It results that optimally 
squeezed channels are robust against signal-mixing, and largely improve 
the strategy power by comparison with coherent ones.\end{abstract}
}
\vspace{-10pt}
\section{Introduction}
The ultimate capacity of a communication network is essentially 
quantum-limited, and the main concern of quantum communication is how to 
discriminate among quantum states that encode the relevant 
information \cite{Chefles}. 
Quantum coding states are generally nonorthogonal, such that they 
cannot be unambiguously discriminated. As a consequence, the detection 
strategy should be optimized at the receiving side, in order 
to maximize the detection probability and/or minimize the transmission 
errors. \par
The scheme we have in mind is the following: a binary alphabet
${\cal A}=\{0,1\}$ with equal {\em a priori} probability symbols is 
being transmitted through a quantum communication channel. The 
information is encoded in two arbitrary Gaussian quantum states 
$\varrho_0$ and $\varrho_1$. In the following we first consider the 
case of pure states $|\psi_0\rangle$ and $|\psi_1\rangle$, whereas, 
in the second part of this letter, the analysis will be extended to 
the mixed-state case. Information is amplitude-keyed encoded \cite{equi}, 
such that the wave functions of the two states are given by
\begin{eqnarray}
\psi_0(x)=\langle x|\psi_0\rangle &=& \frac1{\sqrt{2 \pi \sigma^2}} 
\exp\left[-\frac{x^2}{2\sigma^2}+ i f_0(x)\right]\nonumber\\
\psi_1(x)=\langle x|\psi_1\rangle &=& \frac1{\sqrt{2 \pi \sigma^2}} 
\exp\left[-\frac{(x-a)^2}{2\sigma^2}+i f_1(x)\right]
\label{psis}\;,
\end{eqnarray}
where $f_j(x), j=1,2$ are arbitrary phases, and $a\in{\mathbb R}^+$. 
Since the two states
have the same {\em a priori} probability of being transmitted, the 
mean total energy traveling through the channel is given by $E_T = a^2/2 + 
(\sigma^2-1/2)^2/\sigma^2$ (measured in unit of $h\tau$, $\tau$ being 
the characteristic time of the physical channel, e.g. the period for a  
bounded system, or the time-length of the wave-packet for a free system). 
The case $\sigma^2=1/2$ corresponds to customary
on-off coherent modulation, whereas for $\sigma^2 < 1/2$
we are dealing with {\em squeezed} states \cite{Stoler}. 
Although squeezing increases 
the total energy introduced into the channel, we will show that it can be 
effectively employed to improve the communication scheme, and
to approach the performances of an ideal channel. 
\begin{figure}[h]
\psfig{file=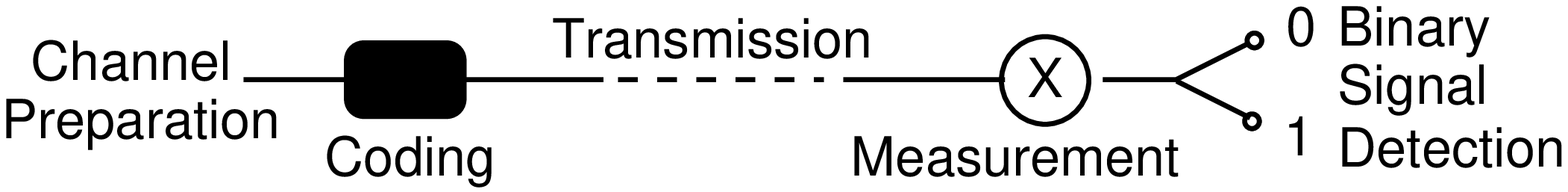,width=80mm}
\caption{Block-diagram of the communication scheme. In the preparation
stage the physical channel is squeezed, then the signal amplitude is applied
or not according to which symbol should be transmitted. At the end of 
the line the signal observable $X$ is measured, and the outcome is 
compared with the threshold value $x_0$ in order to infer which state has
been transmitted. \label{f:cha}}\end{figure}
At the receiver, we consider the standard detection of the signal 
observable $X$, $\hat\mu(x)=|x\rangle\langle x|$, such that the output 
probability densities are given 
by $p_0(x)=|\langle x|\psi_0\rangle|^2=G(x;0,\sigma)$ and $p_1(x;a)=|
\langle x|\psi_1\rangle|^2=G(x;a,\sigma)$, where $G(x;a,\sigma)=(2\pi 
\sigma^2)^{-1/2}\exp[-(x-a)^2/2\sigma^2]$ is a normalized Gaussian of 
mean $a$ and variance $\sigma^2$.
On the basis of each measurement outcome we have to discriminate between 
two hypothesis: the {\em null hypothesis} ${\cal H}_0$ corresponding to the
transmission of $|\psi_0\rangle$ (no signal), and the {\em alternative
hypothesis} ${\cal H}_1$, corresponding to the transmission of 
$|\psi_1\rangle$, {\em i. e.} to the presence of the signal. 
The process of measurement and inference is called a {\em decision strategy}.
We denote by $Q_1$ the {\em power} of the strategy, that is the probability
of inferring the alternative hypothesis when the signal is actually 
present (also called the detection probability), and by $Q_0$ the 
{\em size} of the strategy, {\em i. e.} the probability of inferring 
the alternative hypothesis when the null hypothesis is true 
(also called the false-alarm probability).
\par
In the following we employ a threshold strategy, in which the alternative
hypothesis is chosen if the outcome is greater than a threshold value $x_0$.
In order to determine the threshold value we should optimize the strategy, 
a goal that, in turn, requires to adopt an optimization criterion.
Usually, one uses the criterion of minimizing the average cost of the 
decision, that is, in Bayesian terms, that of minimizing the 
probability of a wrong inference \cite{Helstrom}. 
Alternatively, one may accept to occasionally obtain an inconclusive 
inference in order to achieve error-free discrimination \cite{IDP}.
Actually, these have been fruitful approaches in quantum state 
recognition, especially in the $M$-ary decision problem \cite{YKL}. 
However, the price of a small error probability is usually a 
small detection probability too, which, in turn, may imply the requirement 
of a high repetition rate. On the other hand, in the field of communication 
there exist several protocols that are robust \cite{cov}, 
{\em i.e.} that may satisfactorily work also with a nonzero 
transmission-error rate. In this case, the main interest is that of 
maximizing the detection probability $Q_1$, while maintaining the size 
$Q_0$ to a moderated tolerable level. A decision strategy which is optimized 
according to such a criterion, which we will employ throughout this 
letter, is said to be a Neyman-Pearson (NP) optimized strategy \cite{NP}.
\vspace{-10pt}
\section{Nearly ideal performance of a squeezed channel}
The optimal NP threshold strategy for the present $X$-measurement 
is given in term of a density $\Pi(x)$, which represents the probability 
of choosing the alternative hypothesis after having observed the outcome 
$x$. We have (Neyman-Pearson Lemma) 
\begin{eqnarray}
\Pi (x) = \left\{\begin{array}{ccc}
1&{\rm if}& \Lambda(x) \geq e^\kappa \\
0&{\rm if}& \Lambda(x) < e^\kappa 
\end{array} \right.  \label{npth}\;
\end{eqnarray}
where $\Lambda(x)=p_1(x;a)/p_0(x)$ is the {\em likelihood ratio}, and 
$\kappa$ is the {\em decision level}. By varying the decision level we 
obtain NP strategies with different sizes.
The likelihood ratio is given by $\Lambda(x)=\exp[-(a^2-2ax)/2\sigma^2]$, 
and the NP strategy of Eq.(\ref{npth}) can be summarized as follows: the
alternative hypothesis ${\cal H}_1$ is chosen if the outcome
is greater than the threshold value $x_0=(a^2+2\sigma^2\kappa)/2a$.
The corresponding size and power are given by $Q_0 = \int_{\mathbb R}
\!dx\: \Pi(x)p_0(x)$ and $Q_1= \int_{\mathbb R}\! dx\: \Pi(x)p_1(x;a)$ 
{\em i. e.}
\begin{eqnarray}
Q_0 &=& \int_{x_0}^\infty \!\!\!\! dx\: p_0(x) = \frac12 \left[1-{\rm Erf}
\left(\frac{x_0}{\sigma\sqrt{2}}\right)\right] \label{q0}\;\\
Q_1 &=& \int_{x_0}^\infty \!\!\!\! dx\: p_1(x;a) = 
\frac12 
\left[1-{\rm Erf}\left(\frac{x_0-a}{\sigma\sqrt{2}}\right)\right]
\label{q1}\:,
\end{eqnarray}
By eliminating 
the decision level $\kappa$ between Eqs. (\ref{q0}) and (\ref{q1}) 
one obtains the characteristics $Q_1(Q_0)$
\begin{eqnarray}
Q_1=\frac12 \left\{
1-{\rm Erf} \left[{\rm InvErf}\left(1-2Q_0\right)-\frac{a}{\sqrt{2}\sigma}
\right]\right\}
\label{Q1Char}\;.
\end{eqnarray}
Since the error function ${\rm Erf}(x)$ and its inverse ${\rm InvErf}(x)$ 
are monotone, the power at fixed size increases with the term 
$a/\sqrt{2}\sigma$. As we will see, this quantity may be enhanced 
by squeezing, such that for any energy $E_T$ squeezed 
channels always show larger power than coherent ones. 
\par
\begin{figure}
\psfig{file=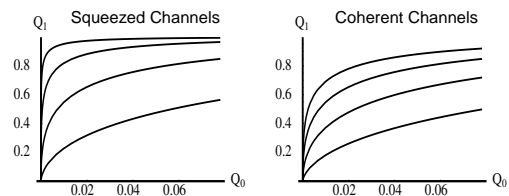,width=7cm}
\caption{Power-size characteristics $Q_1(Q_0)$ of the $X$ measurement 
NP strategy for different channel energies $E_T$. Left: optimally 
squeezed channels. Right: coherent channels. In both plots, from 
bottom to top $E_T=0.5 h\tau, h\tau, 1.5 h \tau, \: {\rm and} \: 
2 h\tau$.\label{f:q0q1}}
\end{figure}

The value of $E_T$ is set by physical constraints, and a question arises 
about the optimal fraction of $E_T$ that should be employed in squeezing 
the channel. In fact, the energy cannot be entirely spent in squeezing, 
since in this case no signal amplitude is left to be discriminated. 
Let us define the squeezing fraction $\gamma$ as the fraction of the 
total energy $E_T$ that is employed to squeeze the channel. In terms of 
$\gamma$ and $E_T$ the amplitude and the squeezing are given by 
$a=\sqrt{2 E_T (1-\gamma)}$ and $\sigma=1/2 (\sqrt{\gamma E_T +2}-
\sqrt{\gamma E_T})$. Using these expressions we have
\begin{eqnarray}
\frac{a}{\sqrt{2}\sigma}= \frac{2\sqrt{E_T(1-\gamma)}}{\sqrt{\gamma 
E_T+2}-\sqrt{\gamma E_T}} \label{a2s}\;.  
\end{eqnarray}
The maximum value is $\left(a/\sqrt{2}\sigma\right)_{max}=
\sqrt{E_T(E_T+2)}$, which is reached for 
\begin{eqnarray}
\gamma_{\sc opt}=\frac12 \frac{E_T}{1+E_T} \label{gotp}\;.  
\end{eqnarray}
$\gamma_{\sc opt}$ thus represents the optimal squeezing fraction to 
discriminate, according to NP criterion, amplitude-keyed signals by 
$X$ measurement. In Fig. \ref{f:q0q1} we show the 
characteristics $Q_1(Q_0)$ for  optimally squeezed and coherent 
channels with different energies. The improvement due to squeezing 
is apparent. 
\par
\begin{figure}
\psfig{file=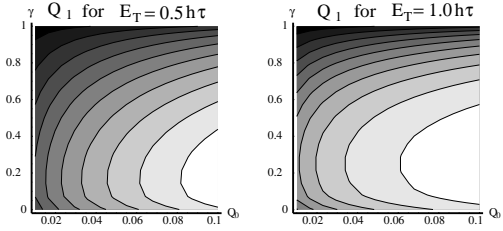,width=7cm}
\caption{Power $Q_1$ of the $X$-measurement NP strategy as a function 
of the size $Q_0$ and the squeezing fraction $\gamma$, for two different
values of the total energy $E_T=0.5 \:h\tau$ and $E_T=1.0\: h\tau$.
\label{f:Qcon}}
\end{figure}

We also notice that $Q_1$ is a smooth function of the squeezing 
fraction, which, in turn, should not be considered as a 
critical parameter. In facts, in order to obtain an enhancement 
of the strategy power, we do not need a fine tuning of $\gamma$.
This is illustrated in Fig. \ref{f:Qcon}, where a contour plot 
of $Q_1$ is shown as a function of $Q_0$ and $\gamma$.
For fixed size $Q_0$ the power slowly varies with $\gamma$ in a 
considerably large range of values.
\par
For a given size $Q_0$ the bound $Q_1 = 1/2$ defines the minimum 
detectable signal. As it follows from Eq. (\ref{Q1Char}), this 
corresponds to $a/\sqrt{2}\sigma={\rm InvErf}(1-2Q_0)$, and using 
Eq. (\ref{a2s}) to
\begin{eqnarray}
E_T^{min}= \frac12 \frac{{\rm InvErf}^2(1-2Q_0)}{1-\gamma +
{\rm InvErf}^2(1-2Q_0) \sqrt{2\gamma(1-\gamma)}}
\label{mdp}\;.
\end{eqnarray}
$E_T^{min}$ decreases with $\gamma$, {\em i.e.} squeezed channels 
allow one to discriminate weaker signals at given size. For 
small $Q_0$, $E_T^{min}$ increases quadratically for coherent 
channels $E_T^{min}=y^2/2$, and only linearly for optimally squeezed 
one $E_T^{min}=-1+\sqrt{1+y^2}$, $y$ being the principal  
solution of the equation $y\sqrt{\pi}Q_0=\exp(-y^2)$. 
\par
In order better to appreciate the benefit of squeezing, we compare 
the power $Q_1$ of the NP $X$-threshold strategy (\ref{npth}), with 
the optimal NP quantum  measurement to discriminate between two 
pure states $|\psi_0\rangle$ and $|\psi_1\rangle$. 
Such an optimal measurement has been found long ago 
\cite{Helstrom,Holevo}, whereas a 
comprehensive approach for mixed states is still 
lacking \cite{Hirota}. For a pair 
of pure states the optimized measurement is given by 
\begin{eqnarray}
\hat\mu (x|\lambda ) =|\psi_1\rangle\langle\psi_1|  
-\lambda|\psi_0\rangle\langle\psi_0| 
\label{opt}\;,
\end{eqnarray}
where $\lambda$ is a Lagrange multiplier which determines the decision 
level. The decision strategy consists in choosing the alternative 
hypothesis ${\cal H}_1$ for positive outcomes, and the resulting 
detection probability reads as follows
\begin{eqnarray}
Q_{1} = \left\{\begin{array}{ll}
\left[\sqrt{Q_{0} \omega} + \sqrt{(1-Q_{0})(1-\omega)}\right]^2
& 0\leq Q_{0} \leq \omega \\ 1 & \omega \leq Q_{0} \leq 1
\end{array}\right. \label{dprob}\;,
\end{eqnarray}
where $\omega=|\langle\psi_0|\psi_1\rangle|^2$ is the overlap 
between the two states. Notice that the ideal NP measurement has 
been considered to find the ultimate quantum limit to high-precision 
binary interferometry \cite{intb}.
\par
\begin{figure}[h]
\psfig{file=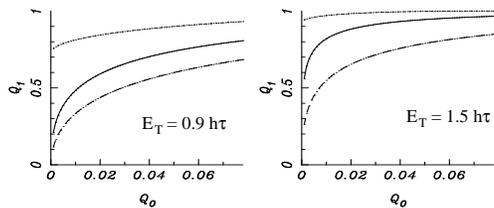,width=65mm}
\caption{Power-size characteristics for two different values 
of the energy $E_T$. Dotted line is the optimal NP strategy, solid
line the (optimally) squeezed channel for $X$-strategy and dashed 
line the coherent one.\label{f:comp}}
\end{figure}

In Fig. \ref{f:comp} we show the power-size characteristics of the 
optimal strategy in comparison with that of the $X$-strategy 
for coherent and optimally squeezed channels. For squeezed channels
the power increases, and approaches the optimal value already for 
moderate energy.
\par
\begin{figure}
\psfig{file=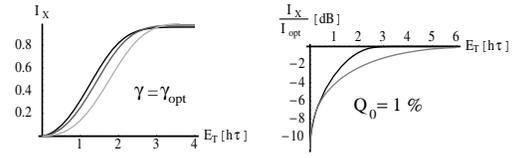,width=7cm}
\caption{Left: mutual information $I_X$ as a function of the 
total energy $E_T$ for $\gamma=\gamma_{\sc opt}$ and for some values of
the size $Q_0=1\%,0.5\%,0.1\%$ (lines in decreasing order of darkness).  
Right: ratio (in dB) between the $X$-strategy mutual information and 
the optimal one at fixed size as a function of the total energy for 
optimally squeezed (black line) and coherent (gray line) channels. 
\label{f:Mit}}
\end{figure}

In order to summarize improvements due to squeezing we consider the mutual 
information between input and output $I=\sum_{ij}P_{ij}p_j\log\left[P_{ij}
/(\sum_j P_{ij}p_j)\right]$, where $p_0=p_1=1/2$ are the {\em a priori} 
probabilities of the two symbols, and $P_{ij}$ is the probability of 
choosing hypothesis ${\cal H}_i$ when hypothesis ${\cal H}_j$ is true. 
In our case, $P_{11}=Q_1$ and $P_{10}=Q_0$, such that $P_{01}=1-Q_1$ 
and $P_{00}=1-Q_0$. On the left panel of Fig. \ref{f:Mit} we show the 
mutual information $I_X$ of the $X$-strategy as a function of the 
total energy $E_T$ for an optimal choice of the squeezing 
fraction $\gamma$. $I_X$ saturates to high value already for moderate
energy, showing only a weak dependence on the size of the strategy.
On the right panel we show the ratio (in dB) between $I_X$ and the
ideal value $I_{opt}$, corresponding to the optimal 
NP strategy. It results that for a squeezed channel the mutual information 
is approaching the ideal value for much lower energy than a coherent one. 
Similar plots are obtained varying the size of the strategies.
\vspace{-10pt}
\section{Fuzzy hypothesis testing and mixed Signals}
So far we have considered information amplitude-keyed on pure states.
However, in practice, it is more likely to deal with mixture, either 
because the coding stage is imperfect, or as a result of noises 
in the transmitter and losses in the channel. For the sake of simplicity, 
we consider a situation in which the null hypothesis still corresponds
to coding onto the vacuum (no amplitude) state, that is $\varrho_0=
|\psi_0\rangle\langle\psi_0|$. On other hand, the alternative hypothesis 
now corresponds to coding the signal on the mixed state
$\varrho_1=\int db\:H_1(b)\:|\psi_b\rangle\langle\psi_b|$, where 
$|\psi_b\rangle$ coincides with $|\psi_1\rangle$ of Eq. (\ref{psis}) 
and $H_1(b)$ is a weight function, which will be taken of Gaussian form.
The two hypothesis to be discriminated are no longer crisp, and 
the decision problem should be formulated in the framework of 
{\em fuzzy hypothesis testing} \cite{Kruse,Arnold}.  
The fuzzy null and alternative 
hypothesis are formulated as follows: ${\cal H}_j$ is true when a 
Gaussian state of amplitude $b$, distributed as $H_j(b)$, is 
transmitted. In our case the two membership density functions are given 
by $H_0(b)=\delta(b)$ and $H_1(b)=(2\pi\Sigma^2)^{-1/2}
\exp[-(b-a)^2/2 \Sigma^2]$. 
\par
In order to analyze the effect of squeezing with a mixed signal we 
need to find the best NP $X$-strategy of its size. Recently, the 
Neyman-Pearson Lemma has been extended to to fuzzy hypothesis 
testing \cite{Taheri}, and this allows us to solve the decision problem.
The NP strategy for mixed states is a density of the form 
(\ref{npth}) with the {\em fuzzy likelihood ratio} given by
\begin{eqnarray}
\label{mixL}\;
\Lambda^{\sc f} (x) &=& \frac{\int db\:p_1(x;b)\:
H_1(b)}{\int db\:p_0(x)\:H_0(b)}=\frac{\int db\:G(x;b,\sigma)\:
H_1(b)}{G(x;0,\sigma)} \\ 
&=& \sqrt{\frac{\beta^2}{1+\beta^2}} \exp \left[
\frac{x^2+2 a x\beta^2-a^2
\beta^2}{2\sigma^2(1+\beta^2)}
\right] \nonumber\:,
\end{eqnarray}
where $\beta^2=\sigma^2/\Sigma^2$. The pure-state case is 
obtained in the limit $\beta\rightarrow\infty$.
\par
The power-size characteristics has the same functional 
form (\ref{Q1Char}) of the pure state case. However, for mixed 
signals part of the energy is degraded to noise, $E_T= (a^2+\Sigma^2)/2 + 
(\sigma^2-1/2)^2/\sigma^2$, such that the amplitude reads as 
follows $a=\sqrt{2E_T(1-\gamma)-\Sigma^2}$. After inserting 
this expression into the term $a/\sqrt{2}\sigma$, and maximizing 
over $\gamma$ one obtains the optimal squeezing fraction for 
mixed channels
\begin{eqnarray}
\gamma_{\sc opt}^{\sc m}=\frac{(2 E_T-\Sigma^2)^2}{8 E_T 
(1+E_T-\Sigma^2/2)}\label{Mgmo}\;.
\end{eqnarray}
As it can be easily proved from Eq. (\ref{Mgmo}) 
$\gamma_{\sc opt}^{\sc m}$ is always smaller then 
$\gamma_{\sc opt}$ for any value of $E_T$, {\em i.e.} 
a smaller amount of squeezing can be employed in a mixed channel 
against a pure channel with the same energy. Correspondingly, 
also the power at fixed size decreases.
However, squeezing a mixed channel is still extremely convenient to 
improve the $X$ strategy by comparison with a mixed coherent 
channel of the same energy. In order to illustrate this behavior, 
we define the ratio $R=I_X^{\sc s}/I_X^{\sc c}$ (at fixed energy 
and mixing parameter) between the mutual information of a 
squeezed and a coherent channel respectively. 
\begin{figure}[h]
\psfig{file=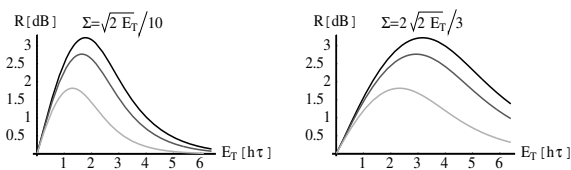,width=8cm}
\caption{
Ratio $R=I_X^{\sc s}/I_X^{\sc c}$ between squeezed and coherent 
mutual information as a function of the total energy $E_T$ for 
a weakly (left) and a strongly (right) mixed channel. 
In both plots curves for different values of the strategy size 
$Q_0$ are shown (black line, $Q_0=0.5\%$, dark gray, $Q_0=1\%$, 
light gray, $Q_0=5\%$).
\label{f:mix}}
\end{figure}\par
In Fig. \ref{f:mix} we show $R$ as a function of $E_T$ for different
values of the size for a weakly ($\Sigma=\sqrt{2 E_T}/10$) and 
a strongly ($\Sigma=2\sqrt{2 E_T}/3$) mixed channel. Notice that 
$\Sigma=\sqrt{2 E_T}$ is the limiting value, corresponding to a 
completely mixed signal with no amplitude and no squeezing.
$R$ linearly increases for small $E_T$ and after a maximum of 
few dB (the actual height depends on the size $Q_0$) 
decreases. In the (unrealistic) limit of very high energies squeezing
the channel is no longer convenient. We notice that for a strongly 
mixed channel such a decreasing is much slower, thus indicating 
that squeezing is effective in a wide range of energies. In other 
words, a squeezed channel is more robust against mixing of signals 
than a coherent one.
\vspace{-10pt}
\section{Conclusions}
In conclusion, we have shown that squeezing the channel in 
amplitude-keyed binary communication increases the detection 
probability at fixed size. We have found the optimal squeezing 
fraction and evaluated the mutual information for both pure 
and mixed signals. Optimally squeezed channels are robust against
signal-mixing, and largely improve the strategy power by comparison
with coherent ones, approaching the performance of the ideal
receiver already for moderate signal energy. 
\vspace{-10pt}
\section*{Acknowledgment}
This work has been cosponsored by C.N.R. and N.A.T.O through  grant No. 217.32 
of the Advanced Fellowship Program. The author thanks people at SESAMO lab 
of the University of Modena and Reggio Emilia for their patience and kind 
hospitality during the last stage of this work.


\begin{thebibliography}{99}
\bibitem{equi} Our results, however, apply also to phase-shift coding, by 
displacing the amplitudes from $(0,a)$ to $(-a/2,a/2)$.
\end{thebibliography}
\end{document}